\documentstyle[11pt]{article}

\input{psfig.sty}

\def\th{\theta} 
\def\pa{\partial}

\def\g{\gamma} 
\def\a{\alpha}
\def\b{\beta}
\def\d{\delta} \def\D{\Delta}
\def\e{\epsilon}

\def\l{\lambda} 
\def\m{\mu}
\def\n{\nu}

\def\s{\sigma} \def\S{\Sigma}

\def\mn{{\mu\nu}}
\def\ab{{\alpha\beta}}
\def\longrightarrow{\relbar\joinrel\relbar\joinrel\rightarrow}%LONG
\def\be{\begin{equation}}
\def\ee{\end{equation}}
\def\bea{\begin{eqnarray}}
\def\eea{\end{eqnarray}}

\begin{document}

\begin{flushright}
BRX TH-524
\end{flushright}

\vspace{.5in}

\begin{center}
{\LARGE\bf Gauged Vector Models and \\

Higher-Spin Representations in AdS$\mbox{\boldmath$_5$}$}

\vspace{.15in}

%\renewcommand{\baselinestretch}{1}
%\small \normalsize

Howard J. Schnitzer\footnote{schnitzr@brandeis.edu}\\
Martin A.\ Fisher School of Physics\\
Brandeis University\\
Waltham, MA 02254

\end{center}

\begin{quotation}
\noindent{\bf Abstract}:  Motivated by the work of Klebanov and
Polyakov [hep-th/020114] on the relationship of the large $N$
O($N$) vector model in three-dimensions to AdS$_4$ and higher spin
representations, we attempt to find analogous connections for
AdS$_5$.  Since the usual O($N$) vector model in four-dimensions
is inconsistent, we consider the (consistent) large $N$ gauged
vector model and a ${\cal N}$=1 supersymmetric analogue in
four-dimensions.  Both these theories have UV and IR fixed points,
and are candidates for a $(\a^\prime )^{-1}$ expansion in AdS$_5$,
a conjectured AdS$_5$/CFT correspondence and higher-spin
representations in the bulk theory.\end{quotation}

%\newpage

\section{Introduction}

The most common and productive application of the AdS$_d$/CFT
correspondence is that of a duality between a conformal field
theory (CFT) on the boundary of the AdS$_d$, and a supergravity
approximation to a string theory \cite{001}.  This is attained in
the large $N$ limit, with $\a^\prime << 1$, which implies that the
AdS radius is large compared to the string scale, and is dual to
the CFT in a strong coupling regime as well.  On the other hand,
one might attempt to interpret any CFT as a theory of quantum
gravity in an asymptotically AdS space-time, with a radius of
curvature of the order of the Planck mass or string scale.  In
terms of string theory, this suggests an AdS/CFT correspondence to
a CFT with an ultraviolet (UV) fixed point, beginning with
$\a^\prime = \infty$, together with $1/\a^\prime$ corrections.
This would correspond to a CFT in a weak coupling regime, with the
expansion around the UV fixed point of the CFT in contrast to the
usual examples of AdS/CFT. More interesting would be examples
where the CFT also had an infra-red (IR) fixed point at which the
theory was also conformal, perhaps even in the perturbative domain
of the CFT. Then one could consider the renormalization group (RG)
flow in the $(\a^\prime)^{-1}$ expansion of the bulk theory from
the UV$\rightarrow$IR as well as in the boundary theory. Such a
study might give new insights into the role of string theory in
the AdS/CFT correspondence.

In this context Klebanov and Polyakov (KP) \cite{002} discussed an
example for AdS$_4$; the large $N$ limit of the O($N$) vector
model in three-dimensions, which has both a UV and IR fixed point,
with the theory being conformal at the IR fixed point.  They
emphasize the advantage of having a CFT with matter in the
fundamental representation of O($N$), rather than the adjoint, as
it makes available for the bulk theory the work of Vasiliev
\cite{003} and others \cite{004,005} on higher-spin
representations.

This paper is an exploratory project aimed at generalizing in an
appropriate way the work of KP to four-dimensions.  The
four-dimensional, large $N$, O($N$) vector model will not work
because it is inconsistent \cite{006}, and parenthetically lacks
an IR fixed point.  A large $N$ gauged vector model coupled to
fermions in the fundamental representation does provide a
consistent model in four-dimensions \cite{007,008}, as do ${\cal
N}$=1 supersymmetric analogues. Both classes of models have both
UV and IR fixed points, so are candidates for consideration of the
kinds of questions raised by KP, but now in four-dimensions.  The
non-supersymmetric gauged vector model \cite{007,008} is presented
in Sec.\ 2. However, it is not known whether this theory is
conformal at the IR fixed point.  An ${\cal N}$=1 cousin of this
model is presented in Sec.\ 3, where supersymmetry ensures
conformal invariance at the IR fixed point. This theory has
Seiberg duality \cite{009}, so that there are both electric and
magnetic descriptions of the theory. Which one is more useful
depends on where one is in the conformal window.

We use the model of Sec.\ 3 to revisit the issues raised by KP,
now in the context of AdS$_5$/CFT.  In Sec.\ 4 we discuss a
possible AdS$_5$/CFT correspondence for the scalar currents of the
theory, and then in Sec.\ 5 extend this to possible higher-spin
representations.  At the IR fixed point, the fundamental fields
have anomalous dimensions, so that a generalization of the
representation theory of Vasiliev  \cite{003} and of Sezgin and
Sundell  \cite{004}, though not yet available, is called for as
one expects the higher-spin currents to have definite but
anomalous dimensions.

Although this work is speculative, it already points to a number
of issues worthy of further study.

\section{The gauged vector model in D=4}

\renewcommand{\theequation}{2.\arabic{equation}}
\setcounter{equation}{0}

The O($N$) vector model in four-dimensions has a long history
\cite{010,006,007}. However, in contrast to the vector model in
three-dimensions, the four-dimensional model is unsuitable for
study of the issues raised by Klebanov--Polykov \cite{002} for (at
least) two reasons; i) the effective potential has no lowest
energy bound as $\phi \rightarrow \infty$, and ii) there is an
absence of an infrared (IR) fixed point.

However, both these difficulties are overcome by the {\it gauged}
U($N$) vector model  \cite{007,008}, coupled to $N_f$ fermions in
the fundamental representation.  This model can be studied in the
large $N$ limit, with $N_f/N$ finite.  [That is, the gauged vector
model is coupled to the Banks--Zak model \cite{011}.]  The
restrictions of asymptotic freedom, and the reality of the
coupling constants throughout the renormalization flows places
important restrictions on $N_f/N$. For massless mesons, these
conditions are sufficiently restrictive to imply the existence of
an IR fixed point $(g_*,\l_*)$ in both gauge and $\l\phi^4$
couplings.  This is a consistent massless theory which is scale
invariant at the IR fixed point, and is in a non-abelian Coulomb
phase.  The Lagrangian density of the massless model is
 \cite{007,008}
\begin{eqnarray}
N^{-1} {\cal L} & = & | \pa_\m \phi + ig
A_\m \phi |^2 + \frac{1}{2\l} \, \chi^2 \nonumber \\
& - &  \chi |\phi |^2 - \frac{1}{4} \, Tr (F_{\m\n} F^{\m\n} ) + i
\sum^{N_f}_{i=1} (\bar{\psi}_i \g \cdot D \psi_i ) \; .
\label{eq:2.1}
\end{eqnarray}
It is important to note that the coupling constants and fields
have been rescaled $(g^2N \rightarrow g^2 ; \; \l N\rightarrow \l
, \; \phi \rightarrow \sqrt{N} \: \phi ; \; A_\m \rightarrow
\sqrt{N} \: A_\m)$  so that $N$ is an overall factor of the
Lagrangian, and hence $N^{-1}$ is a suitable expansion parameter.
Thus in (2.1), $g$ and $\l$ are 't~Hooft couplings. The fields
$\phi$ and $\psi_i$ transform in the fundamental representation of
U(N), the constraint field $\chi$ is a U($N$) singlet, and $D$ the
covariant derivative. Note the absence of a Yukawa coupling
between $\phi$ and $\psi$, as both are in the fundamental
representation.  Study of the model \cite{007,008} indicates that
there is a zero-mass scalar bound state exactly at the IR fixed
point, which appears in $(\phi-\phi )$ scattering. [This state
evolves from the tachyon which is present in the 4d vector model
with $g$=0  \cite{006,010}.]

The renormalized gauge and scalar ('t~Hooft) coupling constants
$g(M)$ and $\l (M)$ depend on an arbitrary mass-scale $M$ as a
result of the renormalization process.  They satisfy the RG eq'ns.
$$
\b_g = M \, \frac{dg}{dM} \eqno{(2.2{\rm a})}
$$
and
$$
\b_\l = M \, \frac{d\l}{dM}  \eqno{(2.2{\rm b})} \label{eq:2.2}
$$
where in the large $N$ limit, with $N_f/N$ fixed is
\cite{011,007,008,012}
\renewcommand{\theequation}{2.\arabic{equation}}
\setcounter{equation}{2}
\begin{eqnarray}
16\pi^2 \b_g & = & -g^3 \left( \frac{22}{3} - \frac{4}{3} \,
\frac{N_f}{N} \right) \nonumber \\[.1in]
& - & \frac{4}{3} \, \frac{g^5}{(4\pi)^2} \, \left( 34-13 \,
\frac{N_f}{N} \right) + \ldots \; , \label{eq:2.3}
\end{eqnarray}
 and
\be \b_\l = a_0 \, \l^2 - a_1\: g^2 \l + a_2\: g^4 \; ,
\label{eq:2.4}
 \ee
  where the $g^2$ dependent coefficients are \cite{007,008,012}
 \be
\left\{
\begin{array}{lcl}
(4\pi )^2 a_0 & = & 2 + 16
\left( \frac{g}{4\pi} \right)^2 + \ldots \nonumber \\[.15in]
(4\pi )^2 a_1 & = & 12 +  \frac{1}{3} \, \left[ 256 - 40 \left(
\frac{N_f}{N} \right)\right]
\left( \frac{g}{4\pi} \right)^2  + \ldots \nonumber \\[.15in]
(4\pi )^2 a_2 & = & 6 +  \frac{1}{3} \, \left[ 304 - 64 \left(
\frac{N_f}{N} \right)\right] \left( \frac{g}{4\pi} \right)^2  +
\ldots
\end{array}
\right. \label{eq:2.5}
 \ee
 Let us review the solution of the coupled RG equations
(2.2)--(2.5) \cite{008}, as the method will be applicable to the
next section.  Since $g(M)$ in (2.3) does not depend on $\l$, one
may solve for it first.  Define
$$
t = ln \, M \eqno{(2.6{\rm a})}\\
$$
$$
x(t) = g^2 (M) \; , \eqno{(2.6{\rm b})}
$$
then from (2.2a)--(2.3)
 $$
  \frac{dx}{dt} = -b_0 \, x^2 + b_1 \, x^3
+ \ldots  \eqno{(2.7{\rm a})}
 $$
If
$$
\frac{34}{13} < \frac{N_f}{N} < \frac{11}{2} \; , \eqno{(2.7{\rm
b})}
$$
then $g^2(M)$ is asymptotically free, and there is an IR
fixed-point for $g^2$, given by \cite{011}
\renewcommand{\theequation}{2.\arabic{equation}}
\setcounter{equation}{7}
 \be \left( \frac{g_*}{4\pi} \right)^2 =
\frac{~~~\left( \frac{11}{2} - \frac{N_f}{N}\right)}
     {13 \left( \frac{N_f}{N} - \frac{34}{13} \right)} \; .
\label{eq:2.8}
 \ee
To solve for the flow of $\l$, use the explicit solution for $g^2
(M)$, and make the change of variables
 \be ds = x(t) dt \; .
\label{eq:2.9}
 \ee
Then
 \be
x(s) = \left[ A \: \exp (sb_0) + \frac{b_1}{b_0} \right]^{-1} \; ,
\label{eq:2.10}
 \ee
 where
$$
 A = \left( \frac{1}{x_0} -
\frac{b_1}{b_0} \right)
$$
with
 \be x(s=0) \equiv  x_0  = g^2 (s=0)\; . \label{eq:2.11}
 \ee
The variables $t$ and $s$ are related by
 \be t = \frac{A}{b_0} \,
[ \exp (sb_0)  - 1] + \frac{b_1}{b_0} \, s \; , \label{eq:2.12}
 \ee
so that $t$ versus $s$ is single-valued, where $s\rightarrow
\pm\infty$ when $t\rightarrow \pm\infty$, with integration
constants chosen so that $t=0$ implies $s=0$.

Define
 \be
  y(s) = \l (s)/g^2 (s) \; . \label{eq:2.13}
\ee
 Then the RG equation for $\l$ becomes
 \be \frac{dy}{ds} = a_0
[y-y_+(s)][y-y_-(s)] \label{eq:2.14}
 \ee
  where the $s$-dependent
coefficients $a_0,\; a_1$ and $a_2$ are given by (\ref{eq:2.5}),
$b_0$, $b_1$ by (\ref{eq:2.3}), (2.7a), and
\begin{eqnarray}
y_\pm (s) & = & \left[ \frac{a_1 -b_0 +b_1 \, x(s)}{2a_0}
\right] \nonumber\\
& \pm & \left\{\left[ \frac{a_1 -b_0 +b_1 \, x(s)}{2a_0} \right]^2
- \frac{a_2}{a_0} \right\}^{1/2} > 0 \label{eq:2.15}
\end{eqnarray}
where reality of the coefficients is required for all values of
$s$.  The curve $y_+(s)$ separates the asymptotically free from
the non-asymptotic free phase.  Flows for $y(s) > y_+(s)$ grow in
the UV to $y (s)_{ \stackrel{\longrightarrow}{s\rightarrow
+\infty}} + \infty$, which is not a consistent phase of the model.
Flows for $y(s) < y_-(s)$ evolve in the IR to $y (s)_{
\stackrel{\longrightarrow}{s\rightarrow -\infty}} - \infty$, which
is excluded, as negative couplings are not allowed.  Therefore
$y_-(s) \leq y(s) \leq y_+ (s)$ is required for consistency, which
describes a non-Abelian Coulomb phase of the theory.

The phase boundary near the UV fixed point is
 \be y_+ (\infty )  =
\frac{2}{3} \left(\frac{N_f}{N} - 1 \right) + \left[ \frac{4}{9}
\, \left(\frac{N_f}{N} -1 \right)^2 -3 \right]^{1/2} \; ,
\label{eq:2.16}
 \ee
which gives the upper-bound of the couplings in the UV,
 \be \l/g^2 \leq
\frac{2}{3} \, \left( \frac{N_f}{N} - 1 \right) + \left[
\frac{4}{9} \, \left( \frac{N_f}{N} - 1 \right)^2 -3 \right]^{1/2}
\; . \label{eq:2.17}
 \ee
Reality of the coupling constants in the UV requires that
(\ref{eq:2.17}) be real, which when combined with (2.7b) gives
 \be 3.6 \simeq \left( \frac{3\sqrt{3}}{2}
 + 1 \right)  \leq N_f/N \leq \frac{11}{2} \; ,
\label{eq:2.18}
 \ee
which is more restrictive than (2.7a).

The RG flow described by (\ref{eq:2.13})-(\ref{eq:2.15}) in the
consistent phase of the theory has a UV and IR fixed point in the
large $N$ limit.  As an example, the RG flow for $N_f/N =5$ is
shown in Fig.\ 1, as extracted from ref. \cite{008}.  We emphasize
that the couplings in the figure are 't~Hooft couplings.  Note
that since $N_f/N$ is close to the upper-bound (\ref{eq:2.18}),
these 't~Hooft couplings are small, justifying perturbation theory
in the large $N$ limit.  The RG flow is already present in the
Banks--Zaks model \cite{011}, to which we have coupled the gauged
vector model.  It is not known whether the IR fixed point of our
model is stable under 1/$N$ corrections.

Notice that because $a_2 \neq 0$ in (\ref{eq:2.5}) and
(\ref{eq:2.15}), (which comes from purely two or more gauge boson
exchange in the scalar scattering), the requirement that couplings
be real is a non-trivial constraint on $N_f/N$, as seen in
(\ref{eq:2.17}). This means that the lower phase-boundary is
$y_-(s) \neq 0$, which implies that in the large $N$ limit, the RG
flow is disjoint from the case where $\l \equiv 0$, which has RG
flow along the real axis $[y_-(s) = 0]$ of Figure 1.  Thus, the
$g^4$ contribution to (\ref{eq:2.4}) leads to the discontinuous
behavior for $\l\rightarrow 0$.

An important issue is whether the theory at the IR fixed is
conformal, and not just scale invariant.  This question is
difficult to resolve, and needs additional study.  Since we are
interested in the possible infinite spin representations for
conformal theories, we turn in the next section to a ${\cal N}=1$
supersymmetric ``cousin" of the gauged vector model described in
this section, where dimensions of chiral operators are protected.
If one could show that the theory of this section is conformal at
the IR fixed point, one could analyze it in analogy to Secs. 4 and
5.

\section{Supersymmetric gauged vector model}

Consider ${\cal N}=1$ supersymmetric QCD with gauge group SU(N),
$N_f$ flavors of quarks  $Q^i$ in the fundamental representation,
$\tilde{Q}_{\tilde{\imath}}$ in the anti-fundamental
representation ($i,\tilde{\imath}$ = 1 to $N_f$), and in addition
a massless chiral superfield $\s$ which is a color and flavor
singlet.  The chiral superfields interact by means of the
superpotential
\renewcommand{\theequation}{3.\arabic{equation}}
\setcounter{equation}{0}
 \be
W = \sqrt{\frac{\l}{N}} \; \s \sum^{N_f}_{i=1} Q^i
\tilde{Q}_{\tilde{\imath}} \; . \label{eq:3.1}
 \ee
The coupling in (\ref{eq:3.1}) has been scaled so that it is the
`t~Hooft coupling, with the same convention as Sec. 2.  Thus the
chiral superfield $\s$ plays a role analogous to $\chi$ in
(\ref{eq:2.1}), except that here we keep it as a propagating
degree of freedom.  [If $\s$ were to be integrated out, one would
have a ${\cal N}=1$ theory analogous to that studied in Sec.\ 2].
This model with $\l =0$ was studied extensively by Seiberg
\cite{009}, while for $\l \neq 0$ one encounters some issues
reminiscent of those considered by Leigh and Strassler \cite{013}.
In our discussion we only consider the non-Abelian Coulomb phase
with $3N/2 < N_f < 3N$, which is the conformal window.

\subsection{RG flow}

The RG equations \cite{022,011} in the $N\rightarrow \infty$,
$N/N_f$ finite limit are determined by the $\b$-functions (where
$g$ and $\l$ are 't~Hooft couplings)
 \be \b_g = - \frac{g^3}{16\pi^2} \: \left( 3-
\frac{N_f}{N}\right) + \frac{g^5}{(16\pi^2 )^2} \left[ 4
\left(\frac{N_f}{N}\right) -6 \right] + \ldots \label{eq:3.2}
 \ee
 and \cite{014}
 \bea
 \b_\l & = & \l [\g_\s + 2\g_Q ] \label{eq:3.3} \\
 & = & 2\l \left[ \left(\frac{N_f}{N} \right) \left( \frac{\l}{16\pi^2}\right) - 2
 \left(\frac{g^2}{16\pi^2} \right) + \ldots \right]\label{eq:3.4}
 \eea
where $\g_\s$ and $\g_Q$ are the anomalous dimensions of $\s$ and
$Q$ respectively.  Note, that due to the ${\cal N}=1$
supersymmetry, only the anomalous dimensions determine $\b_\l$, as
there is no vertex renormalization.   In contrast to
(\ref{eq:2.4}) and (\ref{eq:2.5}), we see that $a_2 \equiv 0$,
which has consequences for the RG flow.  Asymptotic freedom of the
non-Abelian Coulomb phase requires
$$
\frac{3N}{2} < N_f < 3N
$$
and
 \be
  0 \leq \l < 2 \left( \frac{N}{N_f} \right) g^2 \; . \label{eq:3.5}
 \ee
 A necessary condition for an IR fixed point, with $\l \neq 0$, is
 $\g_\s + 2\g_Q = 0$ at the fixed point.

The solution of the RG equations (\ref{eq:3.2}) -- (\ref{eq:3.4})
proceeds exactly as described in Sec.\ 2.  There is a non-trivial
IR fixed point for $g^2$, which for $N_f/N$ = 3 -- $\e$, with $\e
<<1$, is
 \be g^2_* = \frac{8\pi^2}{3} \: \e + {\cal O} (\e^2)\; . \label{eq:3.6}
 \ee
If $0 < \l (M) < 2 (N/N_f )g^2 (M)$, the RG flow for $g^2 (M)$
drives the flow for $\l (M)$ as in Sec.\ 2.  Here we have
 \be
 \b_\l = a_0 \, \l^2 - a_1 g^2 \l \label{eq:3.7}
 \ee
 with
 $$
 \left\{ \begin{array}{l}
 (4\pi )^2 a_0 = 2 \left(\frac{N_f}{N} \right) + \ldots\\
(4\pi ^2 ) a_1 = 4 + \ldots\\
(4\pi ^2 ) a_2 \equiv 0
\end{array}
\right.
$$
  Then defining $y(s) = \l (s)/g^2(s)$ as before,
 \bea
 \frac{dy}{ds} & = & a_0 [y - y_+ (s)][y-y_- (s)] \nonumber \\
 & = & a_0 y [y-y_+ (s)]\label{eq:3.8}
 \eea
 where
 $$
 y_+ (s) = \left[ \frac{a_1 - b_0 + b_1 x(s)}{a_0} \right]
$$
 and
 \be
 y_- (s) = 0
 \; . \label{eq:3.9}
 \ee
 Thus, the flow is as in Fig.\ 1, except the lower phase boundary
is $y_- (s) = 0$.  Hence, here $\l$ may be smoothly turned off, in
contrast to the model of Sec.\ 2.  Therefore, we have an IR fixed
point $g^2_*$ and $\l_*$, which, in analogy with the argument of
Seiberg \cite{009}, should extend throughout the range $3N/2 < N_f
< 3N$.

Given such an IR fixed point, one can use the superconformal
algebra to relate exact results for the dimensions $D$ of
operators, with the $R$-symmetry charge $R$.  They satisfy $D\geq
\frac{3}{2} | R|$, with $D = \frac{3}{2} R$ for chiral operators.
The anomaly free global symmetry is
$$
 SU(N_f) \times SU(N_f) \times U(1)_B \times U(1)_R \; ,
$$
where the quark chiral superfields transform as\footnote{Although
we call the chiral superfields quarks, $Q = \phi + \th \psi_a +
\th^2 F$, so that the bosonic component behaves as in the vector
model.}
 \bea
  Q & : & \left( N_f, 1, 1,  \frac{N_f-N}{N_f} \right) \nonumber \\
\tilde{Q} & : & \left( 1, \bar{N}_f, -1,\frac{N_f-N}{N_f}
\right)\label{eq:3.10}
 \eea
and
$$
 \s \;\; : \;\; \left( 1, 1, 0,  \frac{2N}{N_f} \right) \; . ~~~~~~~~
 $$
 This implies that the gauge invariant operator $Q\tilde{Q}$ has
 dimensions
$$
 D(\tilde{Q}Q)  =  3 \left( \frac{N_f - N}{N_f}
 \right)~~~~~~~~~~~~~~~~~~~~
 \eqno{(3.11{\rm a})} \label{eq:3.11}
$$
$$
 =  2 - \e /3 \;\;\; {\rm for} \;\;\; \e << 1 \; . \eqno{(3.11{\rm b})}
$$
Since $W$ at the fixed point has $R=2$,
$$
 D(\s)  =   \frac{3N}{N_f}
 ~~~~~~~~~~~~~~~~~~~~~~~~~~~~~
 \eqno{(3.12{\rm a})} \label{eq:3.12}
$$
$$
 =  1+ \e /3 \;\;\; {\rm for} \;\;\; \e << 1 \; . \eqno{(3.12{\rm b})}
$$
Note that $D(\s ) > 1$ as is required for an interacting gauge
invariant chiral operator.  At the other end of the conformal
window
\renewcommand{\theequation}{3.\arabic{equation}}
\setcounter{equation}{12}
 \be \left\{ \begin{array}{ll}
 D(\tilde{Q}Q) \rightarrow 1 & \nonumber\\
 & ~~~~~~~{\rm as} \;\; N_f/N \rightarrow 3/2 \; . \\
 D(\s ) \rightarrow 2 \nonumber \end{array}
 \right.  \label{eq:3.13}
 \ee

\subsection{Seiberg duality}

Section 3.1 describes the theory in electric variables.  Following
Seiberg \cite{009}, one also expects a dual magnetic description
of the same theory, even when $\l \neq 0$, where again there is an
IR fixed point in these variables.  The dual group is $SU(\hat{N})
= SU (N_f - N)$, with $N_f$ flavors of quarks $q$ and $\tilde{q}$,
which are not elementary, but non-polynomial functions of $Q$ and
$\tilde{Q}$, with the magnetic description more natural in the
range $3 N/2 \leq N_f < 2N$. In addition there is a gauge singlet
meson $M^j_i$ which appears in the dual description, as well as
the dual of $\s$, denoted by $\hat{\s}$. The assignments of the
quantum numbers of the global symmetry group are
 \bea
 q & : & \left( \bar{N}_f, \: 1, \: \frac{N}{N_f-N} \:,\:
 \frac{N}{N_f}\right)
 \nonumber \\
\tilde{q} & : & \left( 1, \: N_f, \:  \frac{-N}{N_f-N} \:,\:
\frac{N}{N_f} \right)
 \nonumber \\
 M & : & \left( N_f , \: \bar{N}_f , \: 0 , \: 2
 \left(\frac{N_f-N}{N_f} \right)\right) \\
 \hat{\s} & : & \left( 1 , \: 1, \: 0, \: 2
 \left(\frac{N_f-N}{N_f} \right)\right) \; . \nonumber \label{eq:3.14}
 \eea
The gauge singlets have the superpotentials
 $$ W = \sqrt{\frac{\hat{\l}}{\hat{N}}} \; \hat{\s}
 \sum^{N_f}_{i=1} \: \tilde{q}^{\tilde{\imath}}q_i
 $$
 and
\be
 W^\prime = \sqrt{\frac{\l^\prime}{\hat{N}}} \;
M^i_{\tilde{\imath}} \: q_i  \tilde{q}^{\tilde{\imath}}
\label{eq:3.15}
 \ee
where $\hat{N} = (N_f - N)$ and $(\hat{g})^2$, $\hat{\l}$ and
$\l^\prime$ are the `t~Hooft couplings in the magnetic
description.

The $\b$-functions of the magnetic description, in the limit
$N\rightarrow\infty$ and $N_f/N$ fixed, are
 \bea
 \b_{\hat{g}} & = & -\frac{1}{2} \: \left( \frac{\hat{g}^3}
 {16\pi^2} \right) \; \left( \frac{N_f}{N} \right) \nonumber
 \\[.15in]
 & + & \frac{2\hat{g}^5}{(16\pi^2 )^2} \; \left( 3 -
 \frac{N_f}{N}\right) \: \left( 1 + \frac{N_f}{N} \right) + \ldots
 \label{eq:3.16}
 \eea
 and
 \be
 \b_{\hat{\l}} = 2 \hat{\l} \left[ \left(\frac{N_f}{N_f-N}\right)
  \frac{\hat{\l}^2}{(16\pi^2)} -
 2 \left( \frac{\hat{g}^2}{(16\pi^2)}\right) + \ldots \right] \; .
 \label{eq:3.17}
 \ee

 From an argument due to Leigh and Strassler \cite{013}, the flows for
 $\hat{\l}$ and $\l^\prime$ are proportional to each other.
 Therefore the RG flow is entirely analogous to that described in
 Sec.\ 3.1, with asymptotic freedom requiring $\frac{3N}{2} < N_f
 < 3N$ and $0 < \hat{\l} < 2 (1-N/N_f) \hat{g}^2$.  The RG flow may again be
 visualized as in Fig.\ 1, with the lower phase-boundary $y_-(s)
 = 0$.  If $\frac{N_f}{N} = \frac{3}{2} + \hat{\e}$, with
 $\hat{\e} << 1$, the flow in $\hat{g}$ and $\hat{\l}$ is in the
 domain of perturbation theory.  Thus, the addition of (\ref{eq:3.1})
 to the theory does not break Seiberg duality.

 \section{AdS$_5$ and gauged vector models}

 Klebanov and Polyakov (KP) \cite{002} suggested a general relation between
 theories with an infinite number of higher-spin massless gauge
 fields in AdS$_{d+1}$ and large $N$ conformal theories in
 $d$-dimensions containing $N$-component fields with an infinite
 number of conserved currents.  In particular they focused on the
 singlet sector of the 3-d O(N) vector model, and proposed that in
 the large $N$ limit, the vector model was dual to the minimal
 bosonic theory in AdS$_4$ containing massless gauge fields of
 even spin \cite{002,015}.  This proposal has been generalized by
 extending the
 discussion to ${\cal N}$=1 supersymmetry  \cite{016}, as well as to a
 consistency check of the idea  \cite{017}.

 The study of massless higher-spin theories has evolved over many
 years, beginning with the work of Fronsdal \cite{005}, and of Fradkin and
 Vasiliev \cite{003}.  This program has been generalized in several different
 ways \cite{018,019}.  The ideas of KP involve the holographic duals of massless
 higher-spin theories in AdS$_4$.  Little is known about such
 theories so far, and their representations are in turn simpler
 than those in AdS$_5$.  Therefore our interest in the
 relationship of gauged vector models in 4-dimensions to
 higher-spin massless gauge theories is quite speculative.
 Nevertheless it is interesting to explore these issues, despite
 our concerns.

 Motivated by the work of KP, we wish to find duals to large $N$
 conformal theories containing $N$ component fields rather than
 $N\times N$ matrix fields.  As discussed in the introduction, the
 O(N) invariant vector model in $d=4$ has many difficulties
 \cite{006,010}, as
 well as the absence of an IR fixed point.  For these reasons we
 considered the gauged vector model in Sec.\ 2, and an ${\cal
 N}$=1 extension thereof in Sec.\ 3.  Since we are not assured
 that the non-supersymmetric model is conformally invariant at the
 IR fixed point, we will now confine our discussion to the ${\cal
 N}$=1 example of Sec.\ 3.  If it can eventually be shown that the
 model in Sec.\ 2 is conformal at the IR fixed point, then that
 theory could be analyzed as well by the methods of this section.

\renewcommand{\theequation}{4.\arabic{equation}}
\setcounter{equation}{0}
 The ${\cal N}$=1 theory has a spin-zero current
 $J=Q^i\tilde{Q}_{\tilde{\imath}}$, which is a flavor and color
 singlet.  [This is called spin-zero as its lowest component is
 the bosonic spin-zero ``current".]  The dimension of the current
 is
 \be
 D(J) = 3 \left( 1 - \frac{N}{N_f} \right) \; . \label{eq:4.1}
 \ee
Further in the large $N$ limit, the anomalous dimension $D(J)$ is
known in perturbation theory, {\it i.e.},
 \be
  D(J) = 2- \frac{g^2}{8\pi^2} + {\cal O}(g^4) \; . \label{eq:4.2}
 \ee
At the IR fixed point, for $\e << 1$, one has
 \bea
D_{IR}(J) & = & 2-\frac{g^2_*}{8\pi^2} + {\cal O}(g^4_* )
\nonumber
\\
& = & 2 - \frac{\e}{3} + {\cal O}(\e^2 )
 \label{eq:4.3}
 \eea
 while near the UV fixed point
 \be
 D_{UV}(J) = 2 - \frac{g^2}{8\pi^2} +{\cal O}(g^4 )\; . \label{eq:4.4}
 \ee
Following KP, we conjecture that the correlation functions of the
singlet currents at the conformal UV or IR fixed points may be
obtained from AdS$_5$ from an AdS/CFT prescription.  The
dimensions of scalar fields in AdS$_5$ is
 \be \D_\pm = 2 \pm \sqrt{4+(mL)^2}\label{eq:4.5}
 \ee
where $L$ is the radius of AdS$_5$.  Near the UV(IR) fixed point
we identify
$$
 D_{UV}(J)  =  (\D_-)_{UV}
 $$
and
 \be
 D_{IR}(J) =  (\D_-)_{IR}\label{eq:4.6}
 \ee
which implies, from (\ref{eq:4.3})--(\ref{eq:4.5}), that
$$
 (mL^2)_{UV} = -4 + \left( {g^2}/{8\pi^2} \right)^2 + \ldots
 $$
and
 \be
 (mL^2)_{IR} = -4 + \left( {g^2_*}/{8\pi^2} \right)^2 + \ldots
 \label{eq:4.7}
 \ee
for $\e << 1$.  Further, from the RG flow, one has $0 \leq g^2
\leq g^2_*$, so that
 \be
 0 < | m^2 L^2 |_{IR} < |m^2L^2|_{UV} \; ,
 \label{eq:4.8}
 \ee
 which is consistent with the RG flow in the bulk.
 Suppose that the current $J$ is dual to a scalar field $h$ in
 AdS$_5$ \cite{001}, with action
  \be
  S(h) = \frac{N}{2} \: \int d^5x \sqrt{g} \: [(\pa_\m h)^2 +
  m^2h^2
  + \ldots ]\label{eq:4.9}
 \ee
 where $m^2 = - 4/L^2 + {\cal O}(g^4)$, and $L$ is the AdS$_5$
 radius.  This identification required us to choose the
 $\D_-$ in equations
 (\ref{eq:4.5})--(\ref{eq:4.8}) for the CFT, since the anomalous dimension of
 the operator $D(J)$ is
 negative.  [We do not have an interpretation of $\D_+$ in the CFT analogous
 to that of KP \cite{002}.]  There is another independent singlet scalar
 current present if $\s$ is a dynamical chiral superfield.  Define
 \be
 \S = (\s\s )\label{eq:4.10}
 \ee
 whose dimension at the IR fixed point can be expressed in terms
 of $\D_+$, (\ref{eq:4.5}),  but with a perturbative correction.
 At the IR fixed point of $\b_\l$,
(\ref{eq:3.3})  implies that the
 anomalous dimensions satisfy
\be
 \g_\s + \g_J = 0  \; . \label{eq:4.11}
 \ee
 This means that
 \bea
 D_{IR} (\S ) & = & 2\: D_{IR} (\s ) \; = \;
 2-\g_J + \g_\s \nonumber \\[.15in]
 & = & 2 + \left( \frac{g^2_*}{8\pi^2} \right) +
 \left( \frac{N_f}{N} \right)
 \left( \frac{\l_*}{16\pi^2} \right) + \ldots  \: . \label{eq:4.12}
 \eea
Then (\ref{eq:4.3}) and (\ref{eq:4.5}) give
 \bea
 (\D_+ )_{IR}  & = & 2 + \sqrt{4+(mL)^2} \nonumber \\[.15in]
 & = & 2 + \left( \frac{g^2_*}{8\pi^2} \right) + \ldots\label{eq:4.13}
\eea
 Hence
 \bea
 D_{IR}(\S) & = & (\D_+ )_{IR} + \g_\s \nonumber \\[.15in]
 & = &(\D_+ )_{IR} + \left( \frac{N_f}{N} \right)
  \left( \frac{\l_*}{16\pi^2} \right) + \ldots\label{eq:4.14}
 \eea
 Similarly, near the $UV$ fixed point
 \be
 D_{UV} (\S ) = (\D_+)_{UV} +
\left( \frac{N_f}{N} \right)
 \left( \frac{\l}{16\pi^2} \right) + \ldots\label{eq:4.15}
 \ee
 If one associates $\S$ to a dual field $H$ in AdS$_5$ with mass
 $\m$, satisfying
 \be
 D(\S ) = 2 + \sqrt{4+(\m L)^2}\label{eq:4.16}
 \ee
 then
 \be
 \m^2 = -4/L^2 + {\cal O}(g^4, g^2\l , \l^2)\label{eq:4.17}
 \ee
 and so
 \be \m^2 = m^2 + {\cal O}(g^4, g^2\l , \l^2 )\; . \label{eq:4.18}
 \ee
As in (\ref{eq:4.8}), we also have
$$
0 < | \m^2 L^2 |_{IR} < |\m^2L^2 |_{UV} \; .
$$
consistent with RG flow in the bulk. The total action for the
scalar fields $h$ and $H$ in AdS$_5$ is then conjectured to be
 \be
 S = \frac{N}{2} \int d^5x \: \sqrt{g}\: [(\pa_\m h)^2 + (\pa_\m
 H)^2 + m^2h^2 + \m^2 H^2 + \ldots ] \label{eq:4.19}
 \ee
 where the interaction terms should include mixing between $h$ and
 $H$, as well as self-interactions.

 The discussion of this section has focused on the electric
 description of the theory.  For $N_f/N = 3/2 + \hat{\e}$, with
 $\hat{\e} <<1$, the magnetic description should be more useful.
 Define the scalar superfields for the magnetic description,
$$
 \tilde{J} = q^i\tilde{q}_{\tilde{\imath}}
 $$
and
 \be
 \tilde{\S} = \hat{\s}\hat{\s} \; .
 \label{eq:4.20}
 \ee
 Then
 \bea
 D(\tilde{J}) & = & 3N/N_f \nonumber \\
& = & 2 - \frac{4}{3} \: \hat{\e} + {\cal O} (\hat{\e}^2)
\label{eq:4.21}
 \eea
 and
 \bea
 D(\tilde{\S}) & = & 3(1-N/N_f ) \nonumber \\
& = & 1 + \frac{4}{3} \: \hat{\e} + {\cal O} (\hat{\e}^2) \; .
\label{eq:4.22}
 \eea
In a discussion entirely parallel to the electric description, one
associates the scalar fields $\tilde{h}$ and $\tilde{H}$ in
AdS$_5$ to $\tilde{J}$ and $\tilde{\S}$, with an action of the
form of (\ref{eq:4.19}), involving $\tilde{h}$ and $\tilde{H}$,
and masses $\tilde{m}$ and $\tilde{\m}$, where
 \bea
\tilde{m}^2 & = & -4/L^2 + {\cal O} (\hat{g}^4)\nonumber \\[.1in]
 \tilde{m}^2 & = & \tilde{\m}^2  + {\cal O} (\hat{g}^4,
 \hat{g}^2\hat{\l}, \hat{\l}^2)\; .\label{eq:4.23}
\eea
 If $N_f/N$ is not at either end of the conformal window $3/2 N <
N_f < 3N$, then a perturbative expansion in terms of $(g^2,\l )$
or $(\hat{g}^2, \hat{\l})$ is not likely to be rapidly converging,
so that the description in terms of scalar fields in AdS$_5$ is
likely to involve the omitted interactions in (\ref{eq:4.19}) in
an essential way.  Therefore, it may be difficult to give a
suitable explicit completion of  (\ref{eq:4.19}) valid outside the
perturbative end-points of the conformal window.

\section{Infinite Spin Representations}

\renewcommand{\theequation}{5.\arabic{equation}}
\setcounter{equation}{0}

The theories described in Secs.\ 2 and 3 have a class of U(N) or
SU(N) respectively,  gauge and flavor singlet conserved currents
at the UV fixed point $(\hat{g} = \l = 0)$,
 \be
 J_{(\m_1\ldots\m_s)} = \phi^a \stackrel{\leftrightarrow}{D}
 _{(\m_1\ldots}\stackrel{\leftrightarrow}{D}_{\m_s)}
 \phi_a \; + \; {\rm fermion~terms}\label{eq:5.1}
 \ee
 for each spin $s$, where $D_\m$ is the gauge covariant
 derivative.  The model of Sec.\ 3 has an additional class of
 conserved currents of even spins only,
 \be
 \S_{(\m_1\ldots\m_s)} =
  \chi\pa _{(\m_1\ldots}\pa_{\m_s)} \chi\;
  + \; {\rm fermion~terms}\label{eq:5.2}
 \ee
where in (\ref{eq:5.2}) $\chi$ is the bosonic component of the
chiral superfield $\s$.  Following closely KP \cite{002}, and
earlier workers \cite{018}, one conjectures that the correlation
functions of these singlet currents in the free 4-d theories can
be obtained from a bulk action in AdS$_5$, through the  AdS/CFT
property with relates the boundary values of fields with that of
sources $h^{(\m_1\ldots\m_s)}$ in the dual field theory.  That is
\be
 \left< \exp \int d^4x \: h^{(\m_1\ldots\m_s)}_0 J_{(\m_1\ldots\m_s)}
 \right> = e^{S[h_0]}\label{eq:5.3}
 \ee
 where $S[h_0]$ is the action of a high-spin gauge theory in
 AdS$_5$, given in terms of the boundary values $h_0$ of fields.
 [For spin zero, this action is given by (\ref{eq:4.19}).]

 In order to understand the action for the bulk fields
 $h_{(\m_1\ldots\m_s)}$ one needs an appropriate representation
 theory for an infinite tower of higher-spin
 massless-representations in AdS$_5$.  For our purposes, the work of Sezgin
 and Sundell (SS) \cite{004} seems most suitable.  Representations of ${\cal
 N}$=1 supermultiplets in AdS$_5$ are classified by SU(2,2$|$1),
 while the truncation to the bosonic components of the currents
(\ref{eq:5.1}) and (\ref{eq:5.2}) transform according to SO(4,2),
whose group theory has been analyzed by SS \cite{004}.

Denote the bosonic components of the chiral superfields $Q$ and
$\s$ of Sec.\ 3 as $\phi$ and $\chi$ respectively.  At the UV
fixed point, {\it i.e.}, free field limit, the dimensions of
$\phi$ and $\chi$ are
 \bea
 \D(\phi ) & = & \D(\chi ) = 1 \nonumber \\
 & = & 1 + j \hspace{1in} {\rm with} \; j=0 . \label{eq:5.4}
 \eea
Thus $\phi$ and $\chi$ are a pair of doubleton representations of
SO(4,2) \cite{004,005}, denoted by $D_\phi$(0,0;1) and
$D_\chi$(0,0;1) respectively.  As SS \cite{004}  show, one obtains
states with even spins $s=j_L + j_R$ = 0,2,4,$\ldots$ from the
symmetric tensor product of spin-zero doubleton representations,
and odd spins from the anti-symmetric product.  The maximal
compact subgroup of SO(4,2) is SU(2)$_L \times$ SU(2)$_R \times$
U(1)$_R$ whose weight spaces are labeled $D(j_L, j_R ; \D )$, with
lowest weight states
 $| j_L, j_R ;\D >$, where $\D = \frac{3}{2}\: R$ for chiral operators.
 Then
 \bea
 \lefteqn{[D(0,0;1)\otimes D(0,0;1)]_S}~~~~~~~~~ \nonumber \\
 & &=  \sum_{s~{\rm even}} D \left(\frac{s}{2}\;,\frac{s}{2}\; ;\;
 s+2 \right) \label{eq:5.5}
\eea
 \bea
 \lefteqn{[D(0,0;1)\otimes D(0,0;1)]_A}~~~~~~~~~ \nonumber \\
 & &=  \sum_{s~{\rm odd}} D \left(\frac{s}{2}\;,\frac{s}{2}\; ;\;
 s+2 \right)\; . \label{eq:5.6}
\eea

We conjecture that the bosonic components of the currents
(\ref{eq:5.1}) and (\ref{eq:5.2}) are classified by
 \be
  D_\phi (0,0;1)\otimes D_\phi(0,0;1)\label{eq:5.7}
 \ee
and
 \be
  [D_\chi (0,0;1)\otimes D_\chi(0,0;1)]_S \label{eq:5.8}
 \ee
respectively, with the gauge and flavor singlets being selected in
(\ref{eq:5.7}).  Since $\chi$ is real, we only keep the symmetric
product of (\ref{eq:5.6}) in constructing  (\ref{eq:5.8}), while
(\ref{eq:5.7}) requires both (\ref{eq:5.5}) and (\ref{eq:5.6}).
Notice that $s=0$, $\D = 2$ in (\ref{eq:5.5}) which implies for
the bulk states $m^2 = \m^2 = -4/L^2$, corresponding to the
spin-zero currents $J$ and $\S$ of Sec.\ 3.

Sezgin and Sundell \cite{004} build their representations from the
four-component SO(4,1) Dirac spinor $y_\a$, and its conjugate
$\bar{y}^\a$.  They then extend this representation to SU(2,2$|$1)
by introducing an additional set of Grassman odd complex
oscillators $\th$, forming a Clifford algebra and an odd
supercharge $Q_\a = y_\a \th$.  This generates the supersymmetric
extension of hs(2,2) to hs(2,2$|$1) by means of the ${\cal N}$=1
version of the product (\ref{eq:5.5}) and (\ref{eq:5.6}).  We do
not present the details here.

The above discussion is applicable to the UV fixed point, where $g
= \l = 0$.  Now consider the gauge singlet currents (\ref{eq:5.1})
and (\ref{eq:5.2}) for $g^2 \neq 0$, $\l \neq 0$ (not at either
the UV or IR fixed point).  The gauge and flavor singlet bosonic
current $J_{(\mn)}$ has a dependence on the gauge field given by
 \bea
 J_{(\mn)} & = & \phi \: \stackrel{\leftrightarrow}{D}_{(\m}
 \stackrel{\leftrightarrow}{D}_{\n )} \phi \nonumber \\
 & = & \phi \: \stackrel{\leftrightarrow}{\pa}_{(\m }
 \stackrel{\leftrightarrow}{\pa}_{\n )} \phi + g^2 \phi A^\a_\m
 A^\b_\n \{ T_\a, T_\b \} \phi \nonumber \\
 & = & \phi \: \stackrel{\leftrightarrow}{\pa}_{(\m }
 \stackrel{\leftrightarrow}{\pa}_{\n )} \phi + \frac{g^2}{N} \phi A^\a_\m
 A^\a_\n \phi + g^2 d_{\a\b\g} A^\a_\m A^\b_\n (\phi T_\g \phi )
 \label{eq:5.9}
 \eea
 where $g^2$ is the 't~Hooft coupling.  In (\ref{eq:5.9}) $T_\a$
 is a generator of the gauge group SU(N) in the fundamental
 representation, satisfying
 $$
 [T_\a , T_\b ] = i f_{\ab\g}\: T_\g
 $$
 and
 \be
  \{ T_\a , \: T_\b \} = \frac{1}{N} \; \d_{\ab} + d_{\ab\g}
 T_\g \; . \label{eq:5.10}
 \ee
 This can be easily generalized to all of the currents of
 (\ref{eq:5.1}) where the current (\ref{eq:5.9}) and its
generalization to higher spins is defined with normal ordered
fields.  Therefore, the gauge field contribution to the singlet
part of the current is suppressed by
 a factor of $1/N$.  [Recall the rescaling of fields in
 (\ref{eq:2.1}), and in Sec.\ 3.]  Thus, neglect of the gauge
 fields in the gauge singlet, flavor singlet part of the
  currents appears to
 be a consistent truncation, correct to leading order in $1/N$.

There are a very large number of other operators in the CFT
involving gauge field strengths, which are gauge and flavor
singlets. Among these are
\be
 tr F^\a_{(\m_1} \stackrel{\leftrightarrow}{D}_{\m_2}\ldots
 \stackrel{\leftrightarrow}{D}_{\m_{s-1}} F_{\m_s)}{_\a} \label{eq:5.11new}
 \ee
 which have dimensions $\D = 2+s \: (\D \geq 4)$.  Another example is
 \be
 (F^2) = F^{\ab}_{\mn} \: F^{\ab}_{\mn} \label{eq:5.11}
 \ee
 which is dual to the dilation, which has $\D$=4 and AdS$_5$ mass
 $M^2=0$.  There is also a double-trace operator
 \be
 (F^2J) = F^{\ab}_{\mn} \: F^{\ab}_{\mn} J + {\rm fermi \; terms} \; ,
 \label{eq:5.12}
 \ee
with dimensions $\D(F^2J) = 6$ at the UV fixed point,
corresponding to $M^2 = 4/L^2$.  There are infinite number which
generalize
 (\ref{eq:5.11}) or (\ref{eq:5.12}) involving additional field
 strengths.  Other operators are typified by
 \be
 d_{\ab\g} F^{\ab}_{\mn} F^{\b\a}_{\mn} (\phi T_\g \phi ) +
{\rm fermi \; terms} \; , \label{eq:5.13}
 \ee
with D=6, but are not gauge singlets in the matter fields, so are
omitted in our truncation to gauge, flavor singlets.  Similar
remarks apply to the higher-spin fields analogous to
(\ref{eq:5.1}), {\it i.e.}, operators such as $tr (F^2) J_{(\m_1
\ldots \m_s )}$, etc.  Operators
(\ref{eq:5.11new})--(\ref{eq:5.13}) and their generalizations
cannot be represented by (\ref{eq:5.5}) or  (\ref{eq:5.6}).  In
order to include them one will need to consider appropriate
generalizations of the representation theory. This is work for the
future.

 If one is not at the UV fixed point, the currents
 (\ref{eq:5.1}) and  (\ref{eq:5.2}) are not conserved.  Rather
 their divergence can be expressed as a power series in $g^2$ and
 $\l$, with $\l$ related to $g^2$ on the RG flow.  This suggest a
 Higgs-like mechanism for the bulk gauge fields
 $h_{(\m_1\ldots\m_s)}$.  For even spins, the currents (\ref{eq:5.1})
 and  (\ref{eq:5.2}) will mix, as will their bulk gauge field
 duals.  The details of the relevant Higgs mechanism remains to be
 worked out, and is an unsolved problem.

At the IR fixed point the theory is superconformal.  A generic
superconformal primary satisfies unitarity bounds given in
\cite{020}.  The unitarity thresholds are satisfied by massless
fields, and conserved tensor fields.  However, due to the
anomalous dimensions at the IR fixed point, this is not the case
for the currents (\ref{eq:5.1}) or (\ref{eq:5.2}).  Since the
${\cal N}=1$ currents (\ref{eq:5.1}) and (\ref{eq:5.2}) have
anomalous dimensions at the IR fixed point, they are not
conserved, and are no longer the products of fields which saturate
the unitarity lower-bound. This then opens into question the group
theory of (\ref{eq:5.5}) and (\ref{eq:5.8}), as the currents are
no longer the products of doubleton representations at the IR
fixed point. Perhaps a straightforward generalization of
(\ref{eq:5.5})--(\ref{eq:5.8}) for currents with anomalous
dimensions is possible.  This is a project for the future.

\section{Concluding Remarks}

We have presented the large $N$ gauged vector model in
four-dimensions, and an ${\cal N}$=1 supersymmetric analogue, and
studied the RG flow between the UV and IR fixed points of the
theories.  It was speculated that AdS$_5$ duals to the gauge and
flavor singlet currents of the model at the UV fixed point could
be characterized by the higher-spin representations of a type
proposed by Vasiliev and others \cite{003}, and studied in detail
for AdS$_5$ by Sezgin and Sundell \cite{004}.  The higher-spin
currents at the UV fixed-point are conjectured to be dual to
direct products of doubleton representations of SU(2,2$|$1).  The
infinite spin gauge invariance is broken as one moves away from
the UV fixed point.  The higher-spin currents are no longer
conserved, but are possibly Higgsed, with their divergences
proportional to a power series in the `t~Hooft couplings, as
appropriate to a small radius expansion of AdS$_5$  $(1/\a^\prime
<< 1)$.  At the IR fixed point the theory is conformal, though the
currents have anomalous dimensions. Therefore the representation
theory will require a generalization of the work of SS \cite{004},
which is not available as yet.

The discussion of Secs.\ 4 and 5 is speculative, and requires more
work to clarify the status of those ideas.  Further, we do not
have an understanding from the point of view of dual bulk
representations of the non-singlet currents of the model.  Clearly
additional attention to these issues is required, even if the
overall picture presented here is correct.

\newpage

\noindent{\large\bf Acknowledgements}\\

We would like to thank Antal Jevicki, Albion Lawrence, and Niclas
Wyllard for stimulating discussions.  We have benefitted from the
program of the Simons Institute of SUNY, Stony Brook during August
2003, where part of this paper was written.  We wish to
acknowledge the hospitality of the string group at the Harvard
University physics department, which has been extended to me over
a long period of time.  HJS was supported in part by the DOE under
grant DE-FG02-92ER40706.

\newpage

\begin{figure} [h]
\psfig{figure=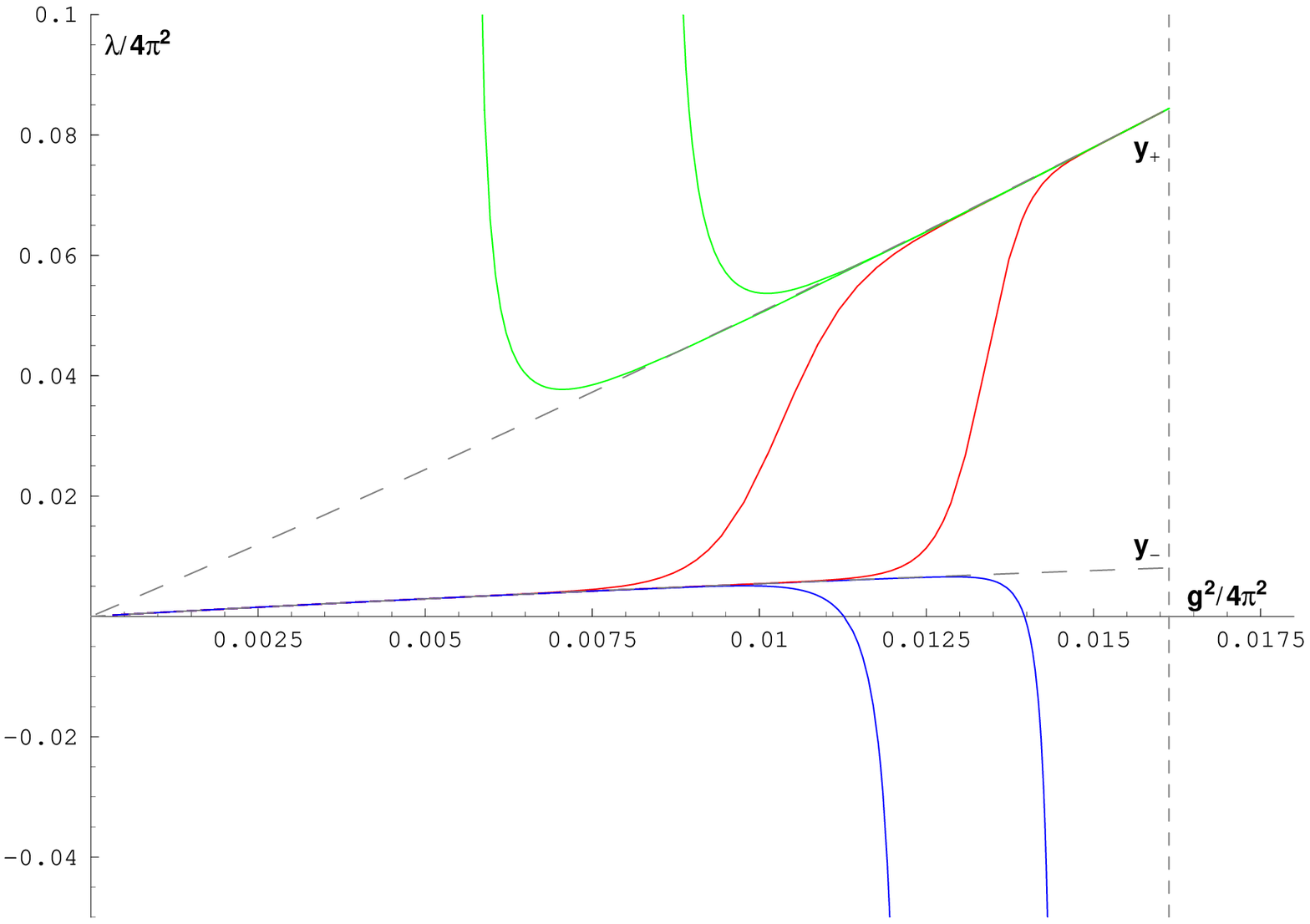,height=15cm,width=15.6cm}
\end{figure}

Graph of the renormalization group flow for $N_f/N=5$, where
ultraviolet to infrared flow progressing from left to right. The
upper and lower dashed lines, $y_+$ and $y_-$ respectively,
brackets flows consistent with asymptotic freedom and stability of
the theory. The vertical dashed line at the right marks the value
of the infrared fixed-point $g_{\star}$.

\end{document}